\documentstyle[frascatiphys,epsfig]{article}

\begin{document}

\title{CP VIOLATION: A THEORETICAL REVIEW  \thanks{Invited talk given at the Second Workshop on Physics and 
Detectors for DAPHNE, Frascati, Italy, April 1995.
To be published in the Workshop Proceedings.}} 

\author{R. D. Peccei\\
 Department of Physics, University of California at Los Angeles,\\
Los Angeles, CA 90095-1547}
\maketitle
\section*{Abstract}
This review of CP violation  focuses on the status of the subject and its likely future development through experiments 
in the Kaon system and with B-decays.  Although present observations of CP violation are
perfectly consistent with the CKM model, we discuss the theoretical
and experimental difficulties which must be faced to establish this
conclusively.  In so doing, theoretical predictions and experimental
prospects for detecting $\Delta S=1$ CP violation through 
measurements of $\epsilon^\prime/\epsilon$ and of rare K decays are
reviewed. The crucial role that B CP-violating experiments will play in elucidating this issue is emphasized .  The importance of looking for evidence for non-CKM 
CP-violating phases, through a search for a non-vanishing transverse
muon polarization in $K_{\mu 3}$ decays, is also stressed.

\section{Introduction}
The discovery more than 30 years ago of the decay $K_{\rm L}\rightarrow 2\pi$
by Christianson, Cronin, Fitch and Turlay \cite{CCFT} provided the
first indication that CP, like parity, was also not a good symmetry 
of nature. It is rather surprising that, for such a mature subject, we have
still so little experimental information available. Indeed, the only firm evidence for CP 
violation to date remains that deduced from meausurements in the neutral Kaon system. 
Here there are five parameters meausured: the values of the the two complex amplitude
ratios for the decays of  $K_{\rm L}$ and $K_{\rm S}$  to $\pi^+\pi^-$ \newline ($n_{+-}=\epsilon+\epsilon^\prime$) and to $\pi^o\pi^o$ ($\eta_{oo}=\epsilon-
2\epsilon^\prime$), plus the semileptonic asymmetry in $K_{\rm L}$ decays ($A_{\rm K_L}$) \footnote{CP Lear \cite{Lear} has presented recently a preliminary  meausurement of the
$K_{\rm S}$ semileptonic asymmetry, $A_{\rm K_S}$, which agrees with $A_{\rm K_L}$ within 10\%.}. However, in fact, these five numbers at present give only one 
{\bf{independent}} piece of dynamical information. To a very good approximation 
\cite{PDG}, $\eta_{+-}$ and $\eta_{oo}$ are equal in magnitude 
\[
|\eta_{+-}|\simeq |\eta_{oo}| \simeq 2\times 10^{-3}~,
\]
and in phase 
\[
\phi_{+-}\simeq \phi_{oo} \simeq 44^{\circ}~,
\]
with the ratio
\[
\epsilon^\prime/\epsilon \leq 10^{-3}~.
\]
Because the two pion intermediate states dominate the neutral Kaon width, assuming 
CPT conservation \cite{DP} the phases $\phi_{+-}$ and $\phi_{oo}$ are 
approximately equal to the superweak phase
\[
\phi_{\rm SW} = \tan^{-1}
\frac{2\Delta m}{\Gamma_{\rm S}-\Gamma_{\rm L}}=
(43.64\pm 0.14)^\circ~. 
\]
CPT conservation also fixes the value of the semileptonic asymmetry in terms of
$ {\rm Re}~ \epsilon$ \cite{Cronin}. Since $\epsilon^{\prime}$ is so small, effectively one has \cite{DP}
\[
A_{\rm K_L} \simeq 2Re~ \eta_{+-}.
\] 
Thus the dynamical information we have today is, essentially, that obtained in the
original discovery experiment\cite {CCFT}, augmented by the statement that there is 
little or no $\Delta S=1$ CP violation!

The above statement is a bit of an exaggeration, since in the last 30 years we have 
learned very much more about CP violation {\bf{outside}} the neutral Kaon complex. In particular,
very strong bounds have been established for the electric dipole moments
of the neutron and the electron \cite{PDG}
\[
d_e,d_n \leq 10^{-25}~{\rm ecm}~.
\]
Furthermore, we have uncovered the fundamental role that CP violation plays in the Universe,
to help establish the observed matter-antimatter asymmetry \cite{Sakharov}. In 
addition, we have a variety of bounds on a host of other CP violating parameters, like
the amplitude ratios $\eta_{+-o}$ and $\eta_{ooo}$ for $K\rightarrow 3\pi$ decays , or
the transverse muon polarization $<P^{\mu}_{\perp}>$ in $K_{\mu 3}$ decays. These bounds, however, are
too insensitive to provide much dynamical information.

In the modern gauge theory paradigm
CP violation can have one of two possible origins.  Either,
\begin{description}
\item[i)] the full Lagrangian of the theory is CP invariant, but this
symmetry is not preserved by the vacuum state: CP $|0\rangle \not= 
|0\rangle$.  In this case CP is a spontaneously broken 
symmetry \cite{TDLee}.
\end{description}
Or
\begin{description}
\item[ii)] there are terms in the Lagrangian of the theory which
are not invariant under CP transformations.  CP is explicitly broken
by these terms and is no longer a symmetry of the theory.
\end{description}

The first possibility, unfortunately, runs into a potential
cosmological problem\cite{KOZ}.  As the universe cools below a
temperature $T^*$ where spontaneous CP violation occurs, one
expects that domains of different CP should form.  These domains
are separated by walls having a typical surface energy density 
$\sigma\sim T^{*^3}$.  The energy density associated with these walls
dissipates slowly as the universe cools further and eventually
contributes an energy density to the universe at temperature T of order
$\rho_{\rm wall}\sim T^{*^3}T$.  Such an energy density today would
typically exceed the universe closure density by many orders of
magnitude:
\[
\rho_{\rm wall}\sim 10^{-7}\left(\frac{T^*}{\rm TeV}\right)^3
{\rm GeV}^{-4} \gg \rho_{\rm closure} \sim 10^{-46}~{\rm GeV}^{-4}~.
\]
One can avoid this difficulty by imaging that the scale where CP is
spontaneously violated is very high, so that $T^*$ is above the
temperature where inflation occurs.  In this case the problem
disappears, since the domains get inflated anyway.  Nevertheless,
there are still problems since it proves difficult to connect
this high energy spontaneous breaking of CP with observed 
phenomena at low energies.  What emerges, in general, are models
which are quite complex \cite{Barr}, with CP violation being associated
with new interactions much as in the original superweak model of
Wolfenstein\cite{superweak}.

If, on the other hand, CP is explicitly broken the phenomenology of
neutral Kaon CP violation is a quite 
natural result of the standard
model of the electroweak interactions.  There is, however, a requirement
emerging from the demand of renormalizability which bears mentioning.
Namely, if CP is explicitly broken then renormalizability requires that
all the parameters in the Lagrangian which can be complex must be
so.   A corollary of this observation
is that the number of possible CP violating
phases in a theory increases with the complexity of the theory, since there
are then more terms which can have imaginary coefficients.

In this respect, the three generation ($N_g=3$) standard model with
only one Higgs doublet is the simplest possible model, since it
has only one phase.  With just one Higgs doublet,
the Hermiticity of the scalar potential allows
no complex parameters to appear.  If CP
is not a symmetry, complex Yukawa couplings are, however, allowed.
After the breakdown of the $SU(2)\times U(1)$ symmetry, these couplings
produce complex mass matrices.  Going to a physical basis with real
diagonal masses introduces a complex mixing matrix in the charged
currents of the theory.  For the quark sector, this is the famous
Cabibbo-Kobayashi Maskawa (CKM) matrix\cite{CKM}.{\footnote{If the
neutrinos are massless, there is no corresponding matrix in the 
lepton sector since it can be reabsorbed by redefining the neutrino
fields.}}  This $N_g\times N_g$ unitary matrix contains $N_g(N_g-1)/2$
real angles and $N_g(N_g+1)/2$ phases.  However, $2N_g-1$ of these phases
can be rotated away by redefinitions of the quark fields leaving only 
$(N_g-1)(N_g-2)/2$ phases.  Thus for $N_g=3$ the standard model has
only one physical complex phase to describe all CP violating 
phenomena.{\footnote{This is not quite true.  In the SM there is also another 
phase related to the QCD vacuum angle which leads to a CP violating
interaction involving the gluonic field strength and its dual. We return to this point in the next section.}}

If CP is broken explicitly, it follows by the renormalizability
corollary that any extensions of the SM will involve further CP
violating phases.  For instance, if one has two Higgs doublets,
$\Phi_1$ and $\Phi_2$, then the Hermiticity of the scalar potential
no longer forbids the appearance of complex terms like
\[
V = \ldots \mu_{12}\Phi_1^\dagger\Phi_2+\mu_{12}^*\Phi_2^\dagger\Phi_1~.
\]
Indeed,  if one did not include such terms 
the presence of complex
Yukawa couplings would induce such terms at one loop.

\section{CP Violation in the Standard Model: Expectations and Challenges}

The gauge sector of the $SU(3)\times SU(2)\times U(1)$ Standard Model contains no
explicit phases since the gauge fields are in the adjoint representation, which is
real, leading to real gauge couplings. Nevertheless, the nontrivial nature of the
gauge theory vacuum \cite{CDG} introduces a phase structure ($\theta$ vacua \cite{JR})
which allows for the presence of effective CP-violating interactions, involving the
non Abelian gauge field strengths and their duals: 
\[
{\cal{L}}_{\rm CP~viol.}=\theta_{\rm strong} \frac{\alpha_{\rm s }}{8\pi}
F_a^{\mu\nu}\tilde F_{a\mu\nu}~+~\theta_{\rm weak} \frac{\alpha_{2}}{8\pi}
W_a^{\mu\nu}\tilde W_{a\mu\nu}~.
\]
The weak vacuum angle $\theta_{\rm weak}$ is actually irrelevant since the electroweak
theory is chiral and through a chiral rotation this angle can be set to zero \cite{AA}. The phase angle $\theta_{\rm strong}$, on the other hand, is problematic. 
First of all, what contributes physically is not $\theta_{\rm strong}$, since
this angle receives
additional contributions from the weak interaction sector as a result of the chiral rotations
that render the quark mass matrices diagonal. Thus, in reality, the CP-violating efffective
interaction is 
\[
{\cal{L}}_{\rm CP~viol.}={\bar\theta}\frac{\alpha_{\rm s }}{8\pi}
F_a^{\mu\nu}\tilde F_{a\mu\nu},
\]
where ${\bar\theta}=\theta_{\rm strong} +{\rm Arg}~{\rm det~M}$ and ${\rm M}$ is the quark mass matrix. The presence of such an
interaction in the Standard Model gives rise to a large contribution to the neutron electric dipole moment. One has, approximately \cite{BC},
\[
d_n\simeq \frac{e}{M_n}(\frac{m_q}{M_n}){\bar{\theta}}~,
\]
with $m_q$ a typical light quark mass.
Thus, to respect the existing experimental bounds on 
$d_n$ \cite{PDG}, ${\bar{\theta}}$ must be extremely small: 
\[
\bar\theta \leq 10^{-9}-10^{-10}.  
\]
Why this should be so is unclear and constitutes
the strong CP problem \cite{strongCP}.

There are three possible attitudes one can take regarding the strong CP problem:
\begin{description}
\item[i)] One can just ignore this problem altogether. Afterall, ${\bar{\theta}}$ is 
yet another uncalculable
parameter in the Standard Model, no different say than the unexplained ratio
$m_e/m_t \sim 10^{-6}$. So why should one worry about this parameter explicitly?
\item[ii)] One can try to calculate ${\bar{\theta}}$ and thereby "explain" the size 
of the neutron electric dipole moment. To do so, one must imagine that CP 
is spontaneously broken, so that indeed ${\bar{\theta}}$ is a finite calculable 
quantity. However, as was mentioned earlier, then one runs into the domain wall 
problem. Models that avoid this problem and which, in principle, produce a tiny 
calculable ${\bar{\theta}}$ exist. However, they are quite recondite \cite{Barr}
and the price one pays for solving the strong CP problem this way is to introduce considerable
hidden underlying structure beneath the Standard Model.
\item[iii)] One can try to dynamically remove ${\bar{\theta}}$ from the theory. This
is my favorite solution, which I suggested long ago with Helen Quinn \cite{PQ}.  Quinn and I proposed solving the strong CP problem by imagining that the  Standard Model has
 an additional global chiral symmetry. The presence of this, so called, $U(1)_{PQ}$
symmetry allows one to rotate away ${\bar{\theta}}$, much as the chiral nature of the 
$SU(2) \times U(1)$ 
electroweak theory allows one to rotate away  $\theta_{weak}$. However, this solution
also requires that axions exist \cite{WeW} and these elusive particles have yet to be detected \cite{strongCP}! An alternative possibility along this vein is that the Standard Model has a
natural chiral symmetry built in, which removes  ${\bar{\theta}}$ because $m_u=0$
\cite{mzero}.However, this solution appears unlikely, as $m_u=0$ is disfavored by current algebra 
analyses \cite{Leutwyler}.
\end{description}
It is fair to say that there is no clear understanding of what to do about the 
strong CP problem at the moment. My own view is that the existence of this 
unresolved problem is something that should not be ignored. There is a message here
and it may simply be that we do not understand CP violation at all!

In the Standard Model, with one Higgs doublet and three generations of quarks and 
leptons, besides the strong CP phase  ${\bar{\theta}}$ there is only one
 other
CP-violating angle in the theory. This is the combination of phases in the Yukawa
couplings of the quarks to the Higgs doublet which remains after all redefinition
of quark fields, leading to a diagonal mass matrix, are done. This weak CP-violating
angle appears as a phase in the CKM mixing matrix, $V$, which details the coupling of the quarks to the charged 
$W$-bosons:   
\[
{\cal{L}}_{CC}=g_2{\bar{u}_i}\gamma_{\mu}(1-\gamma_5)V_{ij}d_jW^{\mu}~+~h.c.~.
\]
However, one does not really know if the complex phase present in the CKM
matrix is responsible for the CP violating phenomena observed
in the neutral Kaon system.  Indeed, one does not know either whether
there are further phases besides the CKM phase.  Nevertheless, it is
remarkable that, as a result of the hierarchial structure of the
CKM matrix and of other dynamical circumstances, one can {\bf qualitatively}
explain all we know experimentally about CP violation today on the
basis of the CKM picture.

\subsection{Testing the CKM Paradigm}

In what follows, I make use of the CKM matrix in the
parametrization adopted by the PDG\cite{PDG} and expand the three
real angles in the manner suggested by Wolfenstein\cite{Wolf} in 
powers of the sine of the Cabibbo angle $\lambda$.  To order $\lambda^3$
one has then
\[
V=\left|
\begin{array}{ccc}
1-\frac{\lambda^2}{2} & \lambda & A\lambda^3(\rho-i\eta) \\
\lambda & 1-\frac{\lambda^2}{2} & A\lambda^2 \\
A\lambda^3(1-\rho-i\eta) & -A\lambda^2 & 1
\end{array}
\right|
\]
with $A,\rho$ and $\eta$ being parameters one needs to fix from
experiments---with $\eta \not= 0$ signalling CP violation.\footnote{It
is often convenient instead of using $\rho-i\eta$ to write this in terms
of a magnitude and phase: $\rho-i\eta=\sigma e^{-i\delta}$, with $\delta$
being the CP violating CKM phase.}

This matrix, with is hierarchical interfamily structure, naturally accounts for the
three principal pieces of independent information that we have today on CP violation. 
As we discussed earlier, these are:
\begin{description}
\item[i)] The value of the  mass mixing parameter $|\epsilon| \sim 10^{-3}$,
which characterizes the strength of the $K_{\rm L}$ to $K_{\rm S}$ amplitude
ratios.
\item[ii)] The small value of the $\epsilon^\prime$ parameter which typifies direct ($\Delta S=1$) CP violation, with the ratio
 $\epsilon^\prime/\epsilon \leq 10^{-3}$.
\item[iii)] The very strong bounds on the electric dipole moments
of the neutron and the electron, which give  $d_e,d_n \leq 10^{-25}~{\rm ecm}$.
\end{description}
One can ``understand" the above three facts quite simply within the Standard Model and
the CKM paradigm.  In the model the parameter $\epsilon$
is determined by the ratio of the imaginary to the real part of the
box graph of Fig. 1a.  It is easy to check that this ratio is of
order
\[
\epsilon \sim \lambda^4 \sin\delta \sim 10^{-3} \sin\delta~.
\]
That is, $\epsilon$ is naturally small because of the suppression of 
interfamily mixing without requiring the CKM phase $\delta$ to be
small.

The explanation of why  $\epsilon^\prime/ \epsilon$ is small is a bit more dynamical.
Basically, this ratio is suppressed  both because of the $\Delta I=1/2$ rule and
because $\epsilon^{\prime}$ arises through the Penguin diagrams of Fig.1b. These
diagrams involve the emission of virtual gluons (or photons \footnote{The contribution
of the electroweak Penguin diagrams are not suppressed by the $\Delta I=1/2$ rule, but these diagrams are only of $O(\alpha)$, not $O(\alpha_{\rm s})$.}), which are Zweig
suppressed\cite{GW}.  Typically this gives
\[
\frac{\epsilon^\prime}{\epsilon} \sim \frac{1}{20} \cdot
\left[\frac{\alpha_{\rm s}}{12\pi}\ln \frac{m_t^2}{m_c^2}\right]
\sim 10^{-3}~.
\]

Finally, in the CKM model the electric dipole moments are small since
the first nonvanishing contributions\cite{Shabalin} occur at three
loops, as shown in Fig. 1c, leading to the estimate\cite{edm}
\[
d_{\rm q,e} \sim em_{\rm q,e}
\left[\frac{\alpha^2 \alpha_{\rm s}}{\pi^3}\right]
\left[\frac{m_t^2 m_b^2}{M_{\rm W}^6}\right] \lambda^6 \sin
\delta \sim 10^{-32}~{\rm ecm}~.
\]

\begin{figure}[t!]
~\epsfig{file=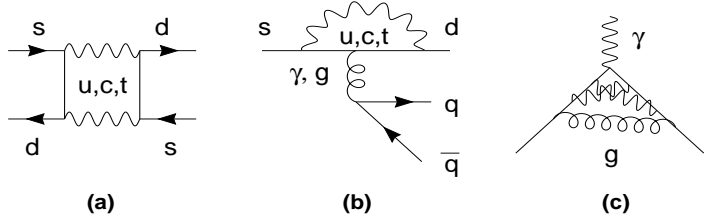,width=13.5cm,height=5cm}
\caption{Graphs contributing to $\epsilon$, $\epsilon^{\prime}$ and $d_{\rm q,e}$}
\end{figure}

One can, of course, use the precise value of $\epsilon$ measured
experimentally to determine an allowed region for the parameters
entering in the CKM matrix.  Because of theoretical uncertainties
in evaluating the hadronic matrix element of the $\Delta S=2$
operator associated with the box graph of Fig. 1a, this parameter
space region is rather large.  Further restrictions on the allowed
values of CKM parameters come from semileptonic B decays and from
$B_d-\bar{B}_{ d}$ mixing.  Because the parameter $A$, related to
$V_{cb}$, is better known, it has become traditional to present the
result of these analyses as a plot in the $\rho-\eta$ plane.  Fig. 2
shows the results of a recent analysis, done in collaboration with my
student, K. Wang\cite{PW}.  The input parameters used, as well as the
range allowed for certain hadronic amplitudes and other CKM matrix
elements is detailed in Table 1

\begin{table}[h!]
\caption[]{
\label{RhoParam}
Parameters used in the $\rho-\eta$ analysis of \cite{PW}
}
\begin{eqnarray*}
\begin{array}{rclr}
|\epsilon| & = & (2.26 \pm 0.02)\times 10^{-3} &\mbox{~~~~~~~\cite{PDG}} \\
\Delta m_d & = & (0.496 \pm 0.032) ps^{-1} & \mbox{\cite{Forty}} \\
m_t & = & (174 \pm 10^{+13}_{-12})~{\rm GeV} & \mbox{\cite{CDF}} \\
|V_{cb}| & = & 0.0378 \pm 0.0026 & \mbox{\cite{Stone}} \\
|V_{ub}|/|V_{cb}| & = & 0.08 \pm 0.02 & \mbox{\cite{Stone}} \\
B_{\rm K} & = & 0.825 \pm 0.035 & \mbox{\cite{Sharpe}} \\
\sqrt{B_d}~f_{B_d} & = & (180 \pm 30)~{\rm MeV} &\mbox{\cite{Lattice}}
\end{array}
\end{eqnarray*}
\end{table}

The resulting $1\sigma$ allowed contour emerging from the overlap
of the three constraints coming from $\epsilon$, $B_{\rm d}-\bar{B}_{\rm  d}$
mixing and the ratio of $|V_{ub}|/|V_{cb}|$, shown in Fig. 3, gives a roughly symmetric
region around $\rho=0$ within the ranges
\[
0.2 \leq \eta \leq 0.5~; \;\; -0.4 \leq \rho \leq 0.4~.
\]

\begin{figure}[t!]
~\epsfig{file=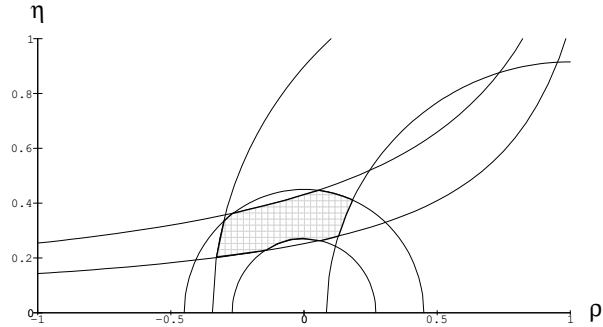,width=13.5cm,height=5cm}
\caption{Constraints on the $(\rho, \, \eta)$ plot}
\end{figure}

As anticipated by our qualitative discussion this region implies
that the CKM phase $\delta$ is large ($\rho=0$ corresponds to
$\delta=\pi/2$).  One should note, however, that this analysis
does not establish the CKM paradigm.  Using only the B physics
constraints one sees that in Fig. 2 there is also an overlap 
region for $\eta=0$, which gives $\rho=-0.33 \pm 0.08$ \cite{PW}.
So one can still imagine that $\epsilon$ is due to some other CP
violating interaction, as in the superweak model \cite{superweak},
with the CKM phase $\delta$ being very small. As Wang
and I \cite{PW} discussed, one may perhaps eliminate this possibility by improving the bounds on $B_{\rm s}-\bar{B}_{\rm s}$ mixing to $\Delta m_s \geq 10 ~{\rm ps}^{-1}$. Since the
present LEP bound from ALEPH is $\Delta m_s \geq 6~{\rm ps}^{-1}$ \cite{ALEPH},
this is not going to be easy. Much more promising, however, is to
try to establish the correctedness of the CKM paradigm by looking at further
tests of CP violation, both in the Kaon system and by meausuring CP-violating
asymmetries in B-decays.

In principle, one can obtain quantitative tests of the CKM model purely with
Kaon experiments.  However, the needed experiments are very challenging,
either due to the high precision required or due to the rarity of the
processes to be studied.  Furthermore, the analysis of these results
is also theoretically very difficult, since it requires
better estimates of hadronic matrix elements than what we have at
present. A good example of both of these challenges is provided by 
$\epsilon^\prime/\epsilon$.  The present data on this ratio is
inconclusive, with the result obtained at Fermilab \cite{E731}:
\[
{\rm Re}~\frac{\epsilon^\prime}{\epsilon}=
(7.4 \pm 5.2 \pm 2.9) \times 10^{-4} ~~~\mbox{[E731]}~,
\]
being consistent with zero within the error, while the result
of the NA31 experiment at CERN \cite{NA31} giving a non-zero value
to $3\sigma$:
\[
{\rm Re}~\frac{\epsilon^\prime}{\epsilon}=
(23.0 \pm 3.6 \pm 5.4) \times 10^{-4} ~~~\mbox{[NA31]}~.
\]
Theoretically, the predictions for $\epsilon^\prime/\epsilon$ are
dependent both on the value of the CKM matrix elements and, more
importantly, on an estimate of certain hadronic matrix elements.

\begin{figure}
~\epsfig{file=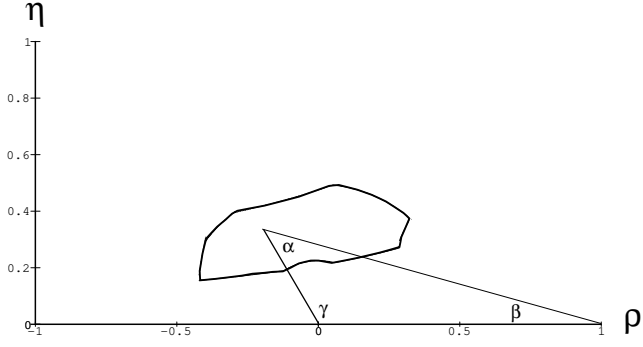,height=5cm}
\caption{Allowed region in the $\rho \, - \, \eta$ plane. Also shown in the plot is a possible unitarity triangle.}
\end{figure}

\subsubsection{Prospects for meausuring $\frac{\epsilon^\prime}{\epsilon}$}

There has been considerable activity recently to try to narrow down the expectations 
for $\epsilon^\prime/\epsilon$ in the Standard Model. To describe the theoretical 
status here, I find it useful to make use of an approximate formula for this ratio 
derived by Buras and Lautenbacher \cite{BL}. These authors express the real
part of this ratio as the sum of two terms \footnote{ Because the difference between
the $I=0$ and $I=2$ $\pi\pi$ phase shifts is also near $45^\circ$ \cite{Gasser}, to
a good approximation $\epsilon^\prime/\epsilon \simeq {\rm Re}~\epsilon^\prime
/\epsilon$.}
\[
{\rm Re}~\frac{\epsilon^\prime}{\epsilon} \simeq
3.6\times 10^{-3} A^2\eta
\left[B_6-0.175\left(\frac{m^2_t}{M_{\rm W}^2}\right)^{0.93}B_8\right]~.
\]
Here $B_6$ and $B_8$ are quantities related to the matrix elements
of the dominant gluonic and electroweak Penguin operators, 
respectively.  The electroweak Penguin contribution is suppressed
relative to the gluonic Penguin contribution by a factor of
$\alpha/\alpha_{\rm s}$.  However, as remarked earlier, it does not suffer from the
$\Delta I=3/2$ suppression and so one gains back a factor of  about 20.
Furthermore, as Flynn and Randall \cite{FR} first noted, the 
contribution of these terms can become significant for large top
mass because it grows approximately as $m_t^2$.  

The result of the CKM analysis presented earlier brackets $A^2 \eta$ in the range
\[
0.12 \leq A^2\eta \leq 0.31~.
\]
For $m_t=175~{\rm GeV}$ the square bracket in the Buras-Lautenbacher formula  
reduces to
[$B_6-0.75B_8$].  Hence one can write the expectation from theory
for $\epsilon^\prime/\epsilon$ as
\[
4.3 \times 10^{-4}[B_6-0.75B_8] \leq
{\rm Re}~\frac{\epsilon^\prime}{\epsilon} \leq 11.2 \times 10^{-4}
[B_6-0.75B_8]~.
\]
Because the top mass is so large, the predicted value for
$\epsilon^\prime/\epsilon$ depends rather crucially on {\bf both} $B_6$
and $B_8$.  These (normalized) matrix elements have been estimated
by both lattice \cite{Ciuchini} and $1/N$ \cite{N} calculations
to be equal to each other, with an individual error of $\pm 20\%$:
\[
B_6=B_8=1 \pm 0.20~.
\]
Thus, unfortunately, the combination entering in $\epsilon^\prime/
\epsilon$ is poorly known.  It appears that the best 
one can say theoretically
is that ${\rm Re}~\epsilon^\prime/\epsilon$ should be a ``few" times
$10^{-4}$, with a ``few" being difficult to pin down more precisely.
Theory, at any rate, seems to favor the E731 experimental result
over that of NA31.

Fortunately, we may learn something more in this area in the near future.
There are 3rd generation experiments in preparation
both at Fermilab (KTeV) and CERN (NA48).  These experiments should
begin taking data in a year or so and are designed to reach statistical
and systematic accuracy for $\epsilon^\prime/\epsilon$ at the level 
of $10^{-4}$.  The Frascati $\Phi$ factory DAPHNE, which should begin 
operations in 1997, in principle, can also provide interesting information
for $\epsilon^\prime/\epsilon$.  At DAPHNE one will need
an integrated luminosity of $\int {\cal{L}}~dt = 10~{\rm fb}^{-1}$ to 
arrive at a statistical sensitivity for $\epsilon^\prime/\epsilon$
at the level of $10^{-4}$.  However, if this statistical sensitivity
is reached, the systematic uncertainties will be quite different than
those at KTeV and NA48, providing a very useful cross-check.  It is
important to remark that, irrespective of detailed theoretical 
predictions, the observation of a non-zero value for $\epsilon^\prime/
\epsilon$ at a significant level would 
provide direct evidence for $\Delta S=1$ CP violation and would rule
out a superweak explanation for the observed CP violation in the 
neutral K sector.

\subsection{Rare Kaon Decays}

There are alternatives to the $\epsilon^\prime/\epsilon$ measurement
which could reveal $\Delta S=1$ (direct) CP violation.  However, these
alternatives involve daunting experiments\cite{RW}, 
which are probably out of reach
in the near term.  Whether these experiments can (or will?) 
eventually be carried out is an open question. Nevertheless, it seems worthwhile here
to try to outline some of the theoretical expectations for these meausurments.

\subsubsection{$K_{\rm S}$ decays}

CP Lear already, and DAPHNE soon, will enable a
more thorough study of $K_{\rm S}$ decays by more efficient
tagging.  The decay $K_{\rm S} \rightarrow 3\pi^o$ is CP-violating, while
the $K_{\rm S}\rightarrow \pi^+\pi^-\pi^o$ mode has both CP-conserving
and CP-violating pieces.  However, even in this case the CP-conserving piece is small and vanishes in the center of the Dalitz
plot.  Hence one can extract information about CP violation from
$K_{\rm S}\rightarrow 3\pi$ decays.  The analogue $K_{\rm S}/K_{\rm L}$
amplitude ratios to $\eta_{+-}$ and $\eta_{oo}$ for $K\rightarrow 3\pi$
have both $\Delta S=1$ and $\Delta S=2$ pieces:
\[
\eta_{ooo}=\epsilon+\epsilon^\prime_{ooo}~; \;\;
\eta_{+-o}=\epsilon+\epsilon^\prime_{+-o}~.
\]
However,
in contrast to what obtains in the $K\rightarrow 2\pi$ case, here
the $\Delta S=1$ pieces can be larger.  Cheng \cite{Cheng} gives
estimates for $\epsilon^\prime_{+-o}/\epsilon$ and
$\epsilon^\prime_{ooo}/\epsilon$ of $O(10^{-2})$, while others
are more pessimistic \cite{pessimistic}.  Even so, there does not
appear to be any realistic prospects in the near future
to probe for $\Delta S=1$
CP-violating amplitudes in $K_{\rm S}\rightarrow 3\pi$.  For instance, at
DAPHNE even with an integrated luminosity of $10~{\rm fb}^{-1}$
one can only reach an accuracy for $\eta_{+-o}$ and $\eta_{ooo}$
of order $3\times 10^{-3}$, which is at the level of $\epsilon$ not
$\epsilon^\prime$.

\subsubsection{Asymmetries in charged K-decays}

CP violating effects involving charged Kaons can only be due to
$\Delta S=1$ transitions, since $K^+\leftrightarrow K^-$ $\Delta S=2$
mixing is forbidden by charge conservation.  A typical CP-violating
effect in charged Kaon decays necessitates a comparison between
$K^+$ and $K^-$ processes.  However, a CP-violating asymmetry between
these processes can occur only if there are at least two decay
amplitudes involved and these amplitudes have {\bf{both}} a relative weak
CP-violating phase and a relative strong rescattering phase between
each other.  Thus the resulting asymmetry necessarily depends on 
strong dynamics. 

 To appreciate this fact, imagine writing the
decay amplitude for $K^+$ decay to a final state $f^+$ as
\[
A(K^+\rightarrow f^+)=A_1~e^{i\delta_{\rm W_1}}e^{i\delta_{\rm S_1}}+
A_2~e^{i\delta_{\rm W_2}}e^{i\delta_{\rm S_2}}~.
\]
Then the corresponding amplitude for the decay $K^-\rightarrow f^-$
is
\[
A(K^-\rightarrow f^-)=A_1~e^{-i\delta_{\rm W_1}}e^{i\delta_{S_1}}+
A_2~e^{-i\delta_{\rm W_2}}e^{i\delta_{\rm S_2}}~.
\]
That is, the strong rescattering phases are the same but one complex
conjugates the weak amplitudes.  From the above, one sees that the
rate asymmetry between these processes is 
\begin{eqnarray*}
{\cal{A}}(f^+;f^-)& = & \frac{\Gamma(K^+\rightarrow f^+)-
\Gamma(K^+\rightarrow f^-)}{\Gamma(K^+\rightarrow f^+)+
\Gamma(K^-\rightarrow f^-)} \\ 
& = & \frac{2A_1A_2\sin(\delta_{\rm W_2}-\delta_{\rm W_1})
\sin(\delta_{\rm S_2}-\delta_{\rm S_1})}
{A_1^2+A_2^2+2A_1A_2\cos(\delta_{\rm W_2}-\delta_{\rm W_1})
\cos(\delta_{\rm S_2}-\delta_{\rm S_1})}~.
\end{eqnarray*}

Unfortunately,
detailed calculations in the standard CKM paradigm for rate
asymmetries and asymmetries in Dalitz plot parameters for various
charged Kaon decays give quite tiny predictions.  This can be
qualitatively understood as follows.  The ratio  
$A_2 \sin(\delta_{\rm W_2}-
\delta_{\rm W_1})/A_1$ is typically that of a Penguin amplitude to
a weak decay amplitude and so is of order $\epsilon^\prime/\epsilon$.
Furthermore, because of the small phase space for
$K\rightarrow 3\pi$ decays, or because one is dealing with electromagnetic
rescattering in $K\rightarrow \pi\pi\gamma$, the rescattering
contribution suppress these asymmetries even more.  Table 2 gives
typical predictions, contrasting them to the expected reach of
the Frascati $\Phi$ factory with $\int {\cal{L}}~dt=10~{\rm fb}^{-1}$.
For the $K\rightarrow 3\pi$ decays, Belkov {\it et al.} \cite{Belkov}
give numbers at least a factor of 10 above those given in Table 2.
However, these numbers are predicated on having very large rescattering
phases which do not appear to be realistic\cite{IMP}.  One is lead to 
conclude that, if the CKM paradigm is correct, it is unlikely that
one will see a CP-violating signal in charged Kaon decays.

\begin{table}
\caption{Predictions for Asymmetries in $K^\pm$ Decays}
\begin{center}
\begin{tabular}{|c|c|c|} \hline
Asymmetry & Prediction & $\Phi$ Factory Reach \\ \hline
${\cal{A}}(\pi^+\pi^+\pi^-;\pi^-\pi^-\pi^+)$ &
$5\times 10^{-8}~~~\mbox{\cite{Pettit}}$ & $3\times 10^{-5}$ \\
${\cal{A}}(\pi^+\pi^o\pi^o;\pi^-\pi^o\pi^o)$  &
$2\times 10^{-7}~~~\mbox{\cite{Pettit}}$ & $5\times 10^{-5}$ \\
${\cal{A}}_{\rm Dalitz}(\pi^+\pi^+\pi^-;\pi^-\pi^+\pi^+)$ &
$2\times 10^{-6}~~~\mbox{\cite{Pettit}}$ & $3\times 10^{-4}$ \\
${\cal{A}}_{\rm Dalitz}(\pi^+\pi^o\pi^o;\pi^-\pi^o\pi^o)$ &
$1\times 10^{-6}~~~\mbox{\cite{Pettit}}$ & $2\times 10^{-4}$ \\  
${\cal{A}}(\pi^+\pi^o\gamma;\pi^-\pi^o\gamma)$ &
$10^{-5}~~~\mbox{\cite{HYC}}$ & $2\times 10^{-3}$ \\  \hline
\end{tabular}
\end{center}
\end{table}

\subsubsection{$K_{\rm L}\rightarrow \pi^o\ell^+\ell^-;~K_{\rm L}
\rightarrow \pi^o\nu \bar\nu$}

Perhaps more promising are decays of the $K_{\rm L}$ to $\pi^o$ plus
lepton pairs.  If the lepton pair is charged, then the process has
a CP conserving piece in which the decay proceeds via a $2\gamma$
intermediate state.  Although there was some initial 
controversy \cite{Seghal}, the rate for the process $K_{\rm L}\rightarrow
\pi^o\ell^+\ell^-$ arising from the CP-conserving $2\gamma$ transitiion
is probably at, or below, the $10^{-12}$ level \cite{Dan}:
\[
B(K_{\rm L}\rightarrow \pi^o\ell^+\ell^-)_{\rm CP~cons.}=
(0.3-1.2)\times 10^{-12}.
\]
Thus this contribution is just a small correction to the dominant CP-violating
amplitude arising from an effective spin 1 virtual state,
$K_{\rm L}\rightarrow \pi^oJ^*$.  Since $\pi^o J^*$ is CP even,
this part of the amplitude is CP-violating. It
has two distinct pieces \cite{Dib}: an
indirect contribution from the CP even piece ($\epsilon K_1$) in the
$K_{\rm L}$ state, and a direct $\Delta S=1$ CP-violating 
piece coming from the $K_2$ part of $K_L$:
\[
A(K_{\rm L}\rightarrow \pi^o J^*)=\epsilon A(K_1\rightarrow \pi^o J^*)+
A(K_2\rightarrow \pi^o J^*)~.
\]

To isolate the interesting direct CP contribution in this process
requires understanding first the size of the indirect contribution.
The amplitude $A(K_1\rightarrow \pi^o J^*)$ could be determined
absolutely if one had a measurement of the process 
$K_{\rm S}\rightarrow \pi^o\ell^+\ell^-$.  Since this is not at
hand, at the moment one has to rely on various guesstimates.
These give the following range for the indirect CP-violating
branching ratio\cite{WW}
\[
B(K_{\rm L}\rightarrow \pi^o\ell^+\ell^-)^{\rm indirect}_{\rm CP~violating}=
(1.6-6)\times 10^{-12}~,
\]
where the smaller number is the estimate coming from chiral
perturbation theory, while the other comes from
relating $A(K_1\rightarrow \pi^o J^*)$ to the analogous amplitude
for charged K decays.

The calculation of the direct CP-violating contribution to the 
process $K_{\rm L}\rightarrow \pi^o\ell^+\ell^-$, as a result of
electroweak Penguin and box contributions and their gluonic
corrections, is perhaps the one that is most reliably known.  The
branching ratio obtained by Buras, Lautenbacher, Misiak and M\"unz in
their next to leading order calculation 
of this process\cite{BLMM} is
\[
B(K_{\rm L}\rightarrow \pi^o\ell^+\ell^-)^{\rm direct}_{\rm CP-violating}=
(5\pm 2)\times 10^{-12}~,
\]
where the error arises mostly from the imperfect knowledge of the CKM
matrix.

Experimentally one has the following 90\% C.L. for the two
$K_{\rm L}\rightarrow \pi^o\ell^+\ell^-$ processes:
\begin{eqnarray*}
B(K_{\rm L}\rightarrow \pi^o\mu^+\mu^-) & < & 5.1\times 10^{-9} \\
B(K_{\rm L}\rightarrow \pi^o e^+e^-) & < & 1.8\times 10^{-9}
\end{eqnarray*}
The first limit comes from the E799 experiment at Fermilab\cite{Harris},
while the second limit combines the bounds obtained by the E845
experiment at Brookhaven\cite{Ohl} and the E799 Fermilab 
experiment\cite{DHH}.  Forthcoming experiments at KEK and Fermilab
should be able to improve these limits by at least an order of
magnitude\footnote{The goal of the KEK 162 experiment is to get to a BR of $O(10^{-10})$ for this mode,
while KTeV hopes to push this BR down to $5\times 10^{-11}$.}, if they
can control the dangerous background arising from the decay
$K_{\rm L}\rightarrow \gamma\gamma e^+e^-$\cite{Greenlee}.  Even more
distant experiments in the future may actually reach the level expected
theoretically for the $K_{\rm L}\rightarrow \pi^o e^+e^-$ 
rate \cite{WINS}.
However, it will be difficult to unravel the direct CP-violating
contribution from the indirect CP-violating contribution, unless
the $K_{\rm S}\rightarrow \pi^oe^+e^-$ rate is also measured
simultaneously.

In this respect, the process $K_{\rm L}\rightarrow \pi^o\nu \bar\nu$
is very much cleaner.  This process is purely CP-violating, since it
has no $2\gamma$ contribution.  Furthermore, it has a tiny indirect CP
contribution, since this is of order $\epsilon$ times the already
small $K^+\rightarrow \pi^+\nu\bar\nu$ amplitude\cite{Littenberg}.
Next to leading QCD calculations for the direct rate have been
carried out by Buchalla and Buras\cite{BBB}, who give the following
approximate formula for the branching ratio for this process
\[
B(K_{\rm L}\rightarrow \pi^o\nu\bar\nu)=8.2\times 10^{-11}
A^4\eta^2\left(\frac{m_t}{M_{\rm W}}\right)^{2.3}~.
\]
This value is very far below the present 90\% C.L. obtained by
the E799 experiment at Fermilab\cite{Weaver}
\[
B(K_{\rm L}\rightarrow \pi^o\nu\bar\nu)<5.8\times 10^{-5}~.
\]
KTeV should be able to lower this bound substantially, perhaps
to the level of $10^{-8}$, but this still leaves a long way to
go!

\subsubsection{$K^+\rightarrow \pi^+\nu\bar\nu$}

The last process I would like to consider in this section is the charged Kaon analogue
to the process just discussed.  Although the decay
$K^+\rightarrow \pi^+\nu\bar\nu$ is not CP violating, it is
sensitive to $|V_{\rm td}|^2 \simeq A^2\lambda^6[(1-\rho)^2+\eta^2]$
and so, indirectly, it is sensitive to the CP violating CKM 
parameter $\eta$.  
For the CP violating decay $K_{\rm L}\rightarrow \pi^o\nu\bar\nu$
the electroweak Penguin and box contributions are dominated by loops
containing top quarks.  Here, because one is not looking at the
imaginary part, one cannot neglect altogether the contribution from charm
quarks.  If one could do so, the branching ratio formula for
$K^+\rightarrow \pi^+\nu\bar\nu$ would be given by an analogous
formula to that for $K_{\rm L}\rightarrow \pi^o\nu\bar\nu$ but with
$\eta^2\rightarrow \eta^2+(1-\rho)^2$.

Because $m_t$ is large, the $K^+\rightarrow \pi^+\nu\bar\nu$ 
branching ratio is not extremely sensitive to the contribution of
the charm-quark loops \cite{Dibc}.  Furthermore, when next to leading
QCD corrections are computed the sensitivity of the result to the
precise value of the charm-quark mass is reduced considerably \cite{BB2}.
Buras {\it et al.}\cite{waiting} give the following approximate
formula for the $K^+\rightarrow \pi^+\nu\bar\nu$ branching ratio
\[
B(K^+\rightarrow \pi^+\nu\bar\nu)=2\times 10^{-11}
A^4\left[\eta^2+\frac{2}{3}(\rho^e-\rho)^2+\frac{1}{3}
(\rho^\tau-\rho)^2\right]
\left(\frac{m_t}{M_{\rm W}}\right)^{2.3}~.
\]
In the above the parameters $\rho^e$ and $\rho^\tau$ differ from
unity because of the presence of the charm-quark contributions.  Taking
$m_t=175~{\rm GeV}$ and $m_c(m_c)=1.30 \pm 0.05~{\rm GeV}$ \cite{GL},
Buras {\it et al.}\cite{waiting} find that $\rho^e$ and $\rho^\tau$
lie in the ranges
\[
1.42 \leq \rho^e \leq 1.55~; \;\; 1.27 \leq \rho^\tau \leq 1.38~.
\]

Using the range of $\eta$ and $\rho$ determined by the CKM 
analysis discussed earlier gives about a 40\% uncertainty for the $K^+\rightarrow \pi^+\nu\bar\nu$
branching ratio, leading to the expectation
\[
B(K^+\rightarrow \pi^+\nu\bar\nu)=(1\pm 0.4)\times 10^{-10}~.
\]
This number is to be compared to the best present limit coming
from the E787 experiment at Brookhaven.  Careful cuts must be made
in the accepted $\pi^+$ range and  momentum to avoid 
potentially dangerous backgrounds, like $K^+\rightarrow \pi^+\pi^o$
and $K^+\rightarrow \mu^+\pi^o\nu$.  There is
a new preliminary result for this branching
ratio \cite{LINS}
\[
B(K^+\rightarrow \pi^+\nu\bar\nu)<3\times 10^{-9}~~~
\mbox{(90\% C.L.)}
\]
which updates the previously published result from the E787 
collaboration \cite{Atiyah}.  This value is still about a factor
of 30 from the interesting CKM model range, but there are hopes
that one can get close to this sensitivity in the present run of
this experiment.

\subsection{The Promise of B-decay CP Violation}

If the CKM paradigm is correct, the analysis of current constraints on the CKM matrix shows that the CP-violating phase $\delta$ is sizable. The reason why one has {\bf small} CP-violating effects in the Kaon sector is solely due to the interfamily mixing suppression. In $B_d$ and $B_s$ decays one can involve all three generations directly and, in certain cases, one can obviate this supression altoghether \cite{BS}. Thus the study of CP violation in B-decays appears very promising.

To produce CP-violating effects, as usual, one has to have interference between two amplitudes that have different CP-violating phases. Because one is interested
in looking for potentially large  sources of CP violation in B-decays, it is important to identify where in the B system sizable phases may reside. Within the CKM paradigm there are two places where large phases appear. The first of these is the relative phase between the $|B_d>$ and $|\bar{B}_d>$ states, which make up the neutral B mass eigenstates:
\[
|B_{d\pm}> \simeq \frac{1}{\sqrt 2}[(1+\epsilon_{B_d})|B_d> \pm (1-\epsilon_{B_d})|\bar{B}_d>]~.
\]
This $ B_d-\bar{B}_d$ mixing phase arises from a box-graph similar to that of Fig. 1a. Here, however, this graph is dominated by the contribution of the top quark loop and one has that
\[
\frac { (1-\epsilon_{B_d})}{ (1+\epsilon_{B_d})} \simeq \frac{V_{td}}{V_{td}^*} =e^{-2i\beta}~,
\]
where $\beta$ is the phase of the $td$ matrix element of the CKM matrix ($V_{td}=|V_{td}|e^{-i\beta}$). If the CKM phase $\delta$ is large so is, in general, $\beta$ since
\[
\tan~\beta=\frac{\sigma \sin~\delta}{1-\sigma \cos~\delta}=\frac{\eta}{1-\rho}~.
\]
The second place where a large phase appears is in any process involving, at the
quark level, a $b \rightarrow u$ transition since $V_{ub}=|V_{ub}|e^{-i\delta}$
provides precisely a meausure of the CKM phase.

The potentially large $B_d-\bar{B}_d$  CP-violating phase $\beta$ is a 
prediction of the CKM paradigm which can be well tested, since this phase is 
unpolluted by strong interaction effects. This is also the case for the 
corresponding mixing phase arising from $B_s-\bar{B}_s$ mixing. However, in 
this case the CKM prediction is that this phase should vanish, since  $V_{ts}$
 is approximately real. So for the case of $B_s$ decays the relevant tests are
  null tests. At any rate, because of $B_d-\bar{B}_d$  mixing, a state 
$|B_{d~{\rm phys}}(t)>$ which at $t=0$ was a pure $|B_d>$
state, evolves in time into a superposition of $|B_d>$ and $|\bar{B}_d> $ 
states:
\[
|B_{d~{\rm phys}}(t)>=e^{-im_{B_d}t}e^{-\Gamma_{B_d}t/2}[\cos\Delta m_{d}t/2~|B_d> 
+ie^{-2i\beta}\sin\Delta m_{d}t/2~|\bar{B}_d>]~.
\]
A similar formula applies for $B_s$ decays. 

It is also possible to isolate cleanly the CP-violating phases 
appearing at the quark level by comparing the decays of $B$ mesons into some
definite final state $f$ to the corresponding 
transition of $\bar{B}$ mesons to the charged-conjugate final state $\bar{f}$.  If these transitions are dominated by a 
single quark decay amplitude \cite{Rosner}, so that
\[
A( B \rightarrow f)=a_fe^{i\delta_f};~~~~~~~A( \bar{B} \rightarrow 
\bar{f})=a_f^*e^{i\delta_f}~,
\]
where $\delta_f$ is a strong rescattering phase, then the ratio of these two 
amplitudes will be  directly sensitive to the quark decay phase. In these 
circumstances, for  decays involving $b\rightarrow u $ transitions, the ratio
\[
\frac{A( \bar{B} \rightarrow \bar{f})}{ A( B \rightarrow f)}\simeq 
\frac{A(b\rightarrow uq\bar{q'})}{A(\bar{b}\rightarrow \bar{u}\bar{q}q')}= 
\frac{V_{ub}}{V_{ub}^*}=e^{-2i\delta}
\]
is a measure of the CP-violating phase $\delta$. On the other hand, the 
corresponding ratio of amplitudes which involve a $b \rightarrow c$ transition at 
the quark level will contain no large CP-violating phase at all, since $V_{cb}$ is real.

In view of the above considerations, the best way to study CP violation in 
neutral $B$-decays is through a comparison of the time evolution of decays of 
states "born" as a $B_d$ (or a $B_s$) into final states $f$, which are CP self-conjugate 
[$\bar{f}=\pm f$], to the corresponding time evolution of states which were born as
 a $\bar{B}_d$ (or a $\bar{B}_s$)  and decay to $\bar{f}$ \cite{Bunk}.  
A straightforward calculation gives for the time dependent rates for these 
processes the expressions:
\[
\Gamma(B_{phys}(t)\rightarrow f)=\Gamma(B \rightarrow 
f)e^{-\Gamma_Bt}[1-\eta_f\lambda_f\sin\Delta m_Bt]
\]
\[
\Gamma(\bar{B}_{phys}(t)\rightarrow \bar{f})=\Gamma(B \rightarrow 
f)e^{-\Gamma_Bt}[1+\eta_f\lambda_f\sin\Delta m_Bt]
\]
In the above $\eta_f$ characterizes the CP parity of the state $f$, 
with $\bar{f}=\eta_f f $ and $\eta_f=\pm 1$, while $\lambda_f$ encapsulates the 
mixing and decay CP violation information for the process. For all decays 
$B\rightarrow f$ which are dominated by just {\bf{one}} weak decay amplitude 
the parameter $\lambda_f$ is free of strong interaction complications  and 
takes one of four values, depending on whether one is dealing with a $B_d$ or 
$B_s$ decay and on whether the decay processes at the quark level involves a $
 b\rightarrow c$ or a $b \rightarrow u$ transition. In these circumstances 
$\lambda_f$ meausures purely CKM information and one finds

\[ \begin{array}{ll} 
\lambda_f=\sin 2\beta & [B_d ~\rm{decays};~b \rightarrow c ~\rm{transition}] \\
\lambda_f=\sin 2(\beta +\delta)\equiv \sin 2\alpha & [B_d ~\rm{decays};~b \rightarrow u ~\rm{transition}] \\
\lambda_f=0 & [B_s ~\rm{decays};~b \rightarrow c ~\rm{transition}] \\
\lambda_f=\sin 2\delta \equiv \sin 2 \gamma & [B_s ~\rm{decays};~b 
\rightarrow u ~ \rm{transition}]~.
\end{array} \]

 The angles $\alpha, \beta$ and $\gamma$ entering in the above equations have 
a very nice geometrical interpretation \cite{bj}. They are the angles of the, 
so called, unitarity triangle in the $\rho-\eta$ plane, whose base is along 
the $\rho$-axis going from $\rho=0$ to $\rho=1$ and whose apex is the point
$(\rho,\eta)$.  That this is the case can be easily deduced by considering the
 $bd$ matrix element of the CKM unitarity equation ($V^{\dagger}V=1$):
\[
V_{ub}^*V_{ud}+V_{cb}^*V_{cd}+V_{tb}^*V_{td}=0~.
\]
To leading order in $\lambda$, the above equation reduces to
\[
V_{ub}^*+V_{td}= A\lambda^3
\]
which, upon dividing by $A\lambda^3$, is precisely the equation describing the
 unitarity triangle.  One possible unitarity triangle, with the angles 
$\alpha,\beta$ and $\gamma$ identified, is shown in Fig. 3. 

Because our present knowledge of the CKM matrix  still allows a considerable 
range for $\rho$ and $\eta$, as shown in Fig. 3, there is considerable 
uncertainty on what to expect for $\sin 2\alpha$, $\sin 2\beta$ and $ \sin 2 
\gamma$. Nevertheless,  from an analysis  of the allowed region in the 
$\rho-\eta$ plane, one can infer the allowed ranges for the unitarity triangle
 angles. In general, one finds that while  both $ \sin 2\alpha$ and $\sin 
2\gamma$ can vanish, $\sin 2 \beta$ is both {\bf{nonvanishing}} and 
{\bf{large}} \cite{Nir}. This is illustrated in Fig. 4, taken from my recent 
examination of this question with Wang \cite{PW}, which plots the presently 
allowed region in the $\sin 2\alpha-\sin 2\beta$ plane. One sees from this 
figure that 
\[
0.23\leq \sin 2 \beta \leq 0.84~.
\]
Thus, in contrast to CP violation phenomena in the Kaon system, for the B 
system within the CKM paradigm there are places where one expects to see large
 effects.

From the above discussion, it is clear that a particularly clean test of the 
CKM paradigm would be provided by the meausurement of $\sin 2\beta$, via the 
observation of a difference in the rates of specific $B_d$ and $\bar{B}_d$ 
decays  to CP self-conjugate states which involve a $b \rightarrow c$ 
transition. A favored mode to study is the decay $B_d \rightarrow \psi K_S$ 
along with  its conjugate \cite{NQ}. This decay has a largish branching ratio \cite{PDG}
 and quite a distinct signature from the leptonic decay of the $\psi$. 
Furthermore, one can argue that this decay is quite clean theoretically. 
Recall that the identification of $\lambda_f$ with one of the angles of the 
unitarity triangle required that the decay rate for the process $B \rightarrow
  f$  be dominated by a single decay amplitude. This, in general, is only 
approximately true. For instance, for the process in question, at the quark 
level both a $ b \rightarrow c$ decay graph {\bf{and}} a $b \rightarrow s$ 
Penguin graph contribute. However,  although this decay involves more than one
 amplitude, both  of these amplitudes have the {\bf{same}} weak decay phase 
\cite{GLP}. The  amplitude involving the quark decay graph has no weak phase 
since it involves a $ b \rightarrow c$ transition. This is also true for the 
$b \rightarrow s$ Penguin graph, since this graph is dominated by the top loop
 contribution which is dominantly real. Hence, effectively, the ratio  of 
$A(\bar{B}_d \rightarrow \psi K_S)$ to $A(B_d \rightarrow \psi K_S)$ is, to a 
very good approximation, unity and $\lambda_{\psi K_S}$ indeed meausures $\sin
 2 \beta$.

Even though $\sin 2\beta$, at least in the CKM paradigm, is large, the 
meausurement of $\lambda_{\psi K_S}$ is far from trivial, since one must be 
able to determine whether the decaying state was born as a $B_d$ or a 
$\bar{B}_d$ and one must have sufficient rate to detect the produced $\psi$ 
though its small leptonic decay mode. Nevertheless, it is clear that future 
meausurements of this and allied decay modes (like $B_d\rightarrow \psi K^* 
\rightarrow \psi K_S \pi $ \cite{KKPS}) perhaps at HERA-B, but certainly at the B factories
 under construction at SLAC and KEK and eventually in hadron colliders, offers
 an excellent chance of verifying- or put into question- the CKM paradigm.

The prospects of meausuring the other two angles of the unitarity triangle, 
$\alpha$ and $\gamma$, and thus of checking the CKM prediction $\alpha + \beta
 +\gamma=\pi$, appear more difficult. These  meausurments are, nevertheless, 
quite important. As Winstein \cite{WEIN} has pointed out, even if one were to 
meausure a large value for $\sin 2\beta$ this still does not totally exclude  
a superweak explanation. One could imagine a perverse superweak model which, 
somehow, contained a small $\Delta S=2$  CP-violating mixing phase, of 
$O(\epsilon)$, but a large $\Delta B=2$  CP-violating mixing phase, $2\beta$. 
 This model can be ruled out by meausuring  the angle $\alpha$ independently 
since, as there are no decay phases, it predicts that $\sin 2\beta=\sin 
2\alpha$.   As Fig. 4 shows, this "superweak'  prediction is not excluded by 
present day data. Thus, if by chance, future meausurements  of $\alpha$ and 
$\beta$ were to fall on the superweak line  one would still need a measurement
 of $\gamma$ to settle the issue \footnote{In superweak models the angle 
$2\gamma $ is the phase associated with $ B_s-\bar{B}_s$   mixing and does not 
necessarily vanish. However, in these models all $B_s$  decays to CP 
self-conjugate states, irrespective of whether they  involve a $b \rightarrow 
u$ or a $b \rightarrow c$ transition, should produce $\lambda_f=\gamma$.}.

\begin{figure}
~\epsfig{file=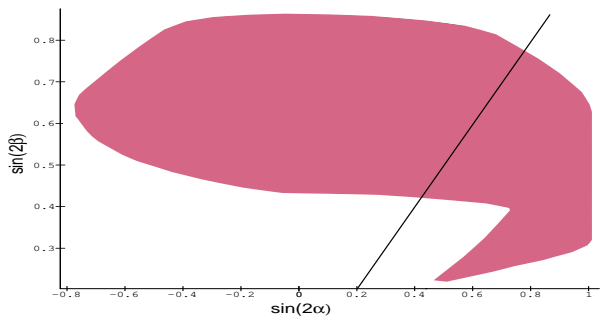,width=12.5cm,height=5cm}
\caption[]{ Plot of the allowed region in the $\sin 2 \alpha -\sin 2 \beta$ 
plane. The vertical line corresponds to the superweak prediction 
$\sin 2 \alpha =\sin 2 \beta$ \cite{WEIN} .}
\end{figure}
 
Most likely, the decay mode $B_d \rightarrow \pi^+\pi^-$ is the process that 
is best suited to study the angle $\alpha$ \cite{NQ}. However, the branching ratio for 
this mode is not yet totally in hand, but is probably quite small, of 
$O(10^{-5})$ \footnote{ Recently, CLEO has observed a few $B_d$ decays into 
pairs of light mesons and has been able to determine the branching ratio for 
the sum of the decay rates into both $\pi K$ and $\pi \pi$ final states: 
$BR(B_d \rightarrow \pi K+\pi\pi) =(1.8 \pm 0.6 \pm 0.2)\times 10^{-5}$ \cite{CLEO}.}. 
This mode also may suffer some Penguin pollution, with estimates ranging from 
1 \% to 10 \% in the amplitude \cite{GLP}.  The CLEO \cite{CLEO}  indications 
that the branching ratio of $B_d\rightarrow \pi K$ is approximately the same 
as that for $B_d \rightarrow \pi \pi$ give one already  some assurances that 
the Penguin amplitude in  $B_d \rightarrow \pi^+\pi^-$ cannot dominate the 
process, since this amplitude is smaller by a factor of $|V_{td}|/|V_{ts}|$ 
compared to the $B_d\rightarrow \pi K$  amplitude. Furthermore, in principle, 
one can isolate the contribution of the Penguin graphs in the process  $B_d 
\rightarrow \pi^+\pi^-$
by meausuring in addition the decays $B_d \rightarrow \pi^o \pi^o$ and 
$B^+\rightarrow \pi^+\pi^o$ \cite{GL2}.  These extra meausurements allow for a 
complete isospin analysis of the amplitudes entering in these two body decays 
and an isolation of the relative rescattering phases. Similar techniques 
\cite{Quinn} may allow the extraction of the angle $\alpha$ in other decay 
modes, like $B_d \rightarrow \rho\pi$, which are not CP self-conjugate.

The determination of the angle $\gamma$ is even more problematic. In 
principle, this angle can be determined by studying the time evolution of 
certain $B_s$ decays. Here, perhaps, the best mode to study would be the decay
 $B_s \rightarrow \pi^o K_S$. However, at the asymmetric B factories now 
under construction, meausurements of $B_s$ decays are unlikely. These 
colliders are optimized to operate at the $\Upsilon(4s)$ and running above the
 $B_s$ production threshold will entail substantial loss of luminosity. Thus a
 determination of $\sin 2\gamma$ from $B_s$ decays will have to await 
dedicated experiments at hadron colliders. One may, however, be able to 
determine $\gamma$, and thus the CKM phase $\delta$, before that by utilizing 
 different techniques. For instance, Gronau and Wyler \cite{GW2}, have shown 
that one could, in principle, extract $\gamma$  by studying the charged B 
decays $B^{\pm}\rightarrow D K^{\pm}$ and their neutral counterparts, using 
isospinology to isolate the rescattering phases. It remains to be seen, 
however, whether this approach can really bear fruit in the presence of 
experimental errors.

\section{Looking for new CP-violating phases}

I would like to argue now  a little more broadly about tests of CP violation. 
Obviously, it is 
very important to check whether the CKM paradigm is correct.  Positive signals
 for 
$\epsilon^\prime/\epsilon \not= 0$ will indicate
the general validity of the CKM picture,  since they require the presence of a
 $ \Delta S=1$
phase.  However, given the large
theoretical uncertainty on the value of this quantity, it is clear that values
 of 
$\epsilon^\prime/\epsilon$ consistent with zero at the $10^{-4}$ level cannot 
disprove
this picture.  In my view, it is  likely that what will
provide the crucial smoking gun for the CKM paradigm are searches 
for CP violation in the B system. The detection of the expected large
asymmetry in $B_d\rightarrow \psi K_{\rm S}$ decays is of paramount 
importance, 
with the meausurement of the other angles in the unitarity triangle and of 
rare Kaon
decays providing eventually a more detailed picture.  However, whether the CKM
picture is (essentially) correct or not,  it is also importat to mount 
experiments 
which may provide the first glimpse at {\bf other} CP-violating
phases, besides the CKM phase $\delta$. Of course, if the CKM picture is 
incorrect then
one knows that at least some of these new phases are superweak in nature, 
arising 
in the $\Delta S=2$ (and, perhaps, in the $\Delta B=2$) sectors. Even if the 
CKM paradigm
is essentially correct, there may be other phases which produce small 
violations in the 
fermion mixing matrix but which are important elsewhere.

Indeed, there are good theoretical arguments for having further CP-violating
phases, besides the CKM phase $\delta$.  For instance, to establish
a matter-antimatter asymmetry in the Universe one needs to have
processes which involve CP violation\cite{Sakharov}.  If the origin
of this asymmetry comes from processes at the GUT scale, then, in
general, the GUT interactions contain further CP-violating phases besides
the CKM phase $\delta~$\cite{PecceiB}.  If this asymmetry is
established at the electroweak scale\cite{Shap}, then most likely
one again needs further phases, both because intrafamily suppression
gives not enough CP violation in the CKM case to generate the asymmetry
and because one needs to
have more than one Higgs doublet\cite{Cohen}.  Indeed this last point
gives the fundamental reason why one should expect to have further
CP-violating phases, besides the CKM phase $\delta$.  It is likely
that the standard model is part of a larger theory.  For instance,
supersymmetric extensions of the SM have been much in vogue
lately.  Any such extensions will introduce further particles and
couplings and thus, by the simple corollary mentioned in the Introduction, 
they will introduce new CP-violating phases.

The best place to look for non-CKM phases is in processes where 
CP violation within the CKM paradigm is either vanishing or very
suppressed.  One such example is provided by experiments aimed at
measuring the electric dipole moments of the neutron or the electron,
since electric dipole moments are predicted to be extremely small
in the CKM model.  Another example concerns measurements of the
transverse muon polarization $\langle P_\perp^\mu\rangle$ in 
$K_{\mu 3}$ decays, which vanishes in the CKM paradigm\cite{Leurer}.
The transverse muon polarization measures a T-violating triple
correlation\cite{Sakurai}
\[
\langle P_\perp^\mu\rangle \sim \langle
\vec s_\mu \cdot (\vec p_\mu\times \vec p_\pi)\rangle~.
\]
In as much as one can produce such an effect also as a result of
final state interactions (FSI) this is not a totally clean test
for new CP-violating phases.  With two charged particles in the
final state, like for the decay $K_{\rm L}\rightarrow \pi^-\mu^+\nu_\mu$,
one expects FSI to give typically
$\langle P_\perp^\mu\rangle_{\rm FSI} \sim \alpha/\pi \sim
10^{-3}$~\cite{Adkins}.  However, for the process
$K^+\rightarrow \pi^o\mu^+\nu_\mu$ with only one charged particle
in the final state, the FSI effects should be much smaller.  Indeed,
Zhitnitski\cite{Zhitnitski} estimates for this proceses that
$\langle P_\perp^\mu\rangle_{\rm FSI}\sim 10^{-6}$.  So
a $\langle P_\perp^\mu\rangle$ measurement in the $K^+\rightarrow
\pi^o\mu^+\nu_\mu$ decay is a good place to test for additional 
CP-violating phases.

The transverse muon polarization $\langle P_\perp^\mu\rangle$ is
particularly sensitive to scalar interactions and thus is a good
probe of CP-violating phases arising from the Higgs 
sector\cite{CFK}.  One can write the effective $K_{\mu 3}$ 
amplitude\cite{BG} as
\[
A=G_{\rm F}\sin\theta_c f_+(q^2)
\left\{p_K^\mu \bar\mu \gamma_\mu(1-\gamma_5)
\nu_\mu + f_S(q^2)m_\mu \bar\mu(1-\gamma_5)\nu_\mu\right\}~.
\]
Then
\[
\langle P_\perp^\mu\rangle = \frac{m_\mu}{M_{\rm K}} {\rm Im}~f_{\rm S}
\left[\frac{|\vec p_\mu|}
{E_\mu+|\vec p_\mu|n_\mu \cdot n_\nu-m_\mu^2/M_{\rm K}}\right]
\simeq 0.2~{\rm Im}~f_{\rm S}~.
\]
Here $n_\nu(n_\nu)$ are unit vectors along the muon (neutrino)
directions and the numerical value represents the expectation after
doing an average over phase space\cite{Kuno}.

\begin{figure}
~\epsfig{file=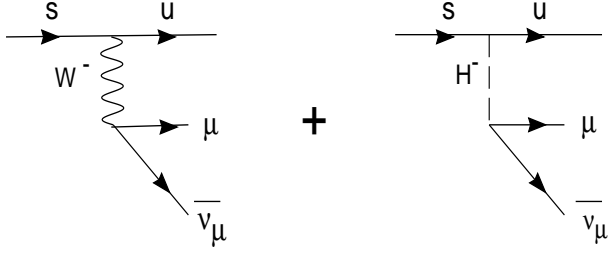,width=13.5cm,height=3.0cm}
\caption{Graphs contributing to $\langle P_\perp^\mu\rangle$}
\end{figure}

Contributions to ${\rm Im}~f_{\rm S}$ can arise in multi-Higgs models
(like the Weinberg 3-Higgs model\cite{Weinberg}) from the charged
Higgs exchange shown in Fig. 5, leading to \cite{Higgs}
\[
{\rm Im}~f_{\rm S} \simeq {\rm Im}(\alpha^*\gamma)
\frac{M_{\rm K}^2}{M_{{\rm H}^-}^2}~.
\]
Here $\alpha(\gamma)$ are constants associated with the charged
Higgs coupling to quarks (leptons).  Because a leptonic vertex
is involved, one in general does not have a strong constraint on
${\rm Im}(\alpha^*\gamma)$.  By examining possible non-standard
contributions to the B semileptonic decay $B\rightarrow X\tau \nu_\tau$,
Grossman\cite{Grossman} obtains
\[
{\rm Im}(\alpha^*\gamma) < 
\frac{0.23~M^2_{H^-}}{[{\rm GeV}]^2}
\]
which yields a bound for $\langle P_\perp^\mu\rangle$ of
$\langle P_\perp^\mu\rangle < 10^{-2}$.  Amusingly, this is
the same bound one infers from the most accurate measurement of
$\langle P_\perp^\mu\rangle$ done at Brookhaven about a decade
ago \cite{Blatt}, which yielded
\[
\langle P_\perp^\mu\rangle = (-3.1\pm 5.3)\times 10^{-3}~.\]
In specific models, however, the leptonic phases associated with
charged Higgs couplings are correlated with the hadronic phases. 
In this case,
one can obtain more specific restrictions on $\langle P_\perp^\mu\rangle$
due to the strong bounds on the neutron electric dipole moment.
For instance, for the Weinberg 3 Higgs model, one relates
${\rm Im} (\alpha^*\gamma)$ to a similar product of couplings of
the charged Higgs to quarks\cite{Higgs}:
\[
{\rm Im}(\alpha^*\gamma)=\left(\frac{v_u}{v_e}\right)^2~
{\rm Im}(\alpha^*\beta)~,
\]
where $v_u~(v_e)$ are the VEV of the Higgs doublets which couples
to up-like quarks (leptons).  The strong bound on the neutron electric
dipole moment\cite{PDG} then gives the constraint
\[
{\rm Im}(\alpha^*\beta) \leq 
\frac{4\times 10^{-3}~M_{{\rm H}^-}^2}{[{\rm GeV}]^2}~.
\]
If one assumes that $v_u \sim v_e$, this latter bound gives
a strong constraint on $\langle P_\perp^\mu\rangle \break 
[\langle P_\perp^\mu\rangle < 10^{-4}]$.  
However, this constraint is removed if
$v_u/v_e \sim m_t/m_\tau$.

Similar results are obtained in the simplest supersymmetric
extension of the SM.  In this case, ${\rm Im}~f_{\rm S}$ arises from
a complex phase associated with the gluino mass.  Assuming all
supersymmetric masses are of the same order, Christova and 
Fabbrichesi\cite{CF} arrive at the estimate
\[
{\rm Im}~f_{\rm S} \simeq \frac{M_{\rm K}^2}{m_{\tilde g}^2}
\frac{\alpha_{\rm s}}{12\pi} \sin\phi_{\rm susy}~,
\]
where $\phi_{\rm susy}$ is the gluino mass CP-violating phase.
This phase, however, is strongly restricted by the neutron electric
dipole moment.  Typically, one finds\cite{Hall}
\[
\sin\phi_{\rm susy} \leq \frac{10^{-7}~m^2_{\tilde g}}
{[{\rm GeV}]^2}
\]
leading to a negligible contribution for
$\langle P_\perp^\mu\rangle$,  below the level of 
$\langle P_\perp^\mu\rangle_{\rm FSI}$.

An experiment (E246) is presently underway at KEK aimed at
improving the bound on $\langle P_\perp^\mu\rangle$ obtained
earlier at Brookhaven.  The sensitivity of E246 is such that one should
be able to achieve an error $\delta\langle P_\perp^\mu\rangle \sim
5\times 10^{-4}$\cite{Kuno}.  This level of precision is very
interesting and, in some ways, it is comparable or better to
$d_n$ measurements for probing CP-violating phases from the scalar
sector.  This is the case, for instance, in the Weinberg model if
$v_u/v_e$ is large.  At any rate, if a positive signal were
to be found, it would be a clear indication for a non-CKM CP-violating
phase.  Furthermore, as Garisto\cite{Garisto} has pointed out, a
positive signal at the level aimed by the E246 experiment would
imply very large effects in the corresponding decays in the B system
involving $\tau$-leptons (processes like $B^+\rightarrow D^o \tau^+
\nu_\tau$), since one expects, roughly,
\[
\langle P_\perp^\tau\rangle_{\rm B} \sim
\frac{M_{\rm B}}{M_{\rm K}} \frac{m_\tau}{m_\mu}
\langle P_\perp^\mu\rangle_{\rm K}~.
\]
Thus, in principle, a very interesting experimental cross-check
could be done.

\section{Concluding Remarks}

I would like to conclude more or less in the way in which I started this 
review, 
by reemphasizing that even after thirty years from the discovery of CP 
violation this 
phenomena remains shrouded in mystery. However, there are some grounds for 
optimism.
It is quite possible that before the year 2000 we shall know whether the CKM 
model 
provides the approximately correct description of CP violation. For instance, 
a convincing
 non zero determination of $\epsilon^{\prime}/\epsilon$ would exclude the 
superweak
hypothesis, while a meausurement of $\sin 2\beta \sim O(1)$ would strongly 
favor the 
CKM explanation. Both of these experimental results could be on hand in this 
time frame.

A more detailed understanding of the full CKM structure, or  a further 
understanding of
CP violation if the CKM paradigm fails, will be more difficult. The 
meausurement of 
$\alpha$ is probably the simplest of the more difficult things to accomplish. 
A 
direct meausurement of the CKM phase $\delta$, or equivalently the angle 
$\gamma$, by
means of  experiments in the B sector or through the study of rare K decays, 
is very
challenging. So are attempts at finding non CKM phases, although experiments 
searching 
for these effects are to be encouraged since they would signal  new physics 
beyond the Standard Model. Indeed, it is important  also that other  CP 
violation
experiments where one expects very small effects in the Standard Model be 
pushed
to their limits, as surprises may arise. A case in point is provided by
searches for CP violation in charged K decays, or in $K_S$ decays, 
to be carried out here at DAPHNE, where little is really known. It is unclear,
 however, 
whether all this experimental activity will be able to throw any light on 
the strong CP problem. Nevertheless, if we are to really understand CP 
violation, one
day we will have to understand why $\bar{\theta}\simeq 0$.  

\section*{Acknowledgments}          
  
Part of the material on CP violation in K-decays presented here is drawn from a talk I gave at the 23rd INS Symposium in Tokyo, Japan \cite{PINS}. This work was supported in part by the
Department of Energy under Grant No. FG03-91ER40662.

\vspace{1pc}
%\section{References}


\begin{thebibliography}{999}
\bibitem{CCFT} J. H. Christenson, J. W. Cronin, V. L. Fitch, and
R. Turlay, Phys. Rev. Lett. {\bf 13}(1964) 138.
\bibitem{Lear}  B. Pagels, to appear in the Proceedings of the 23rd INS Symposium, Tokyo, Japan, 1995.
\bibitem{PDG} Particle Data Group: L. Montanet {\it et al.},
Phys. Rev. D{\bf 50}(1994) 1173.
\bibitem{DP} C. O. Dib and R. D. Peccei, Phys. Rev. D{\bf 46}(1992)
2265.
\bibitem{Cronin} J. W. Cronin, Rev. Mod. Phys. {\bf 53}(1981) 373.
\bibitem{Sakharov} A. Sakharov, JETP Lett. {\bf 5}(1967) 24
\bibitem{TDLee} T. D. Lee, Phys. Reports {\bf 9C}(1974) 148.
\bibitem{KOZ} I. Kobzarev, L. Okun, and Y. Zeldovich, Phys. Lett.
{\bf 50B}(1974) 340.
\bibitem{Barr} S. Barr, Phys. Rev. D{\bf 30}(1984) 1805;
A. Nelson, Phys. Lett. {\bf 143B}(1934) 165.  See also, R. D. Peccei,
in {\bf 25 Years of CP Violation}, ed. J. Tran Than Van (Editions Frontiers,
Gif-sur-Yvette, 1990).
\bibitem{superweak} L. Wolfenstein, Phys. Rev. Lett. {\bf 13}(1964) 562.
\bibitem{CKM} N. Cabibbo, Phys. Rev. Lett. {\bf 10}(1963) 531;
M. Kobayashi and T. Maskawa, Prog. Theor. Phys. {\bf 49}(1973) 652.
\bibitem{CDG} C. G. Callan, R. Dashen and D. Gross, Phys. Lett. {\bf 63B}(1976) 334.
\bibitem{JR} R. Jackiw and C. Rebbi, Phys. Rev. Lett. {\bf 37}(1976) 172. 
\bibitem{AA} N. V. Krasnikov, V. A. Rubakov and V. F. Tokarev, J. Phys {\bf A12}(1979) L343. For a pedagogical discussion see also, A. A. Anselm and A. A. Johansen, Nucl. Phys. {\bf B407}(1993) 652.
\bibitem{BC} V. Baluni, Phys. Rev. {\bf D19}(1979) 2227: R. Crewther, P. Di Vecchia, G. Veneziano and E. Witten, Phys. Lett. {\bf 88B}(1979) 123: {\bf 91B}(1980) 487 (E).
\bibitem{strongCP} For a review, see, for example, R. D. Peccei in
{\bf CP Violation} ed. C. Jarlskog (World Scientific, Singapore, 1989).
\bibitem{PQ} R. D. Peccei and H. R. Quinn, Phys. Rev. Lett. {\bf 38}(1977) 1440; Phys. Rev. {\bf D16}(1977) 1791.
\bibitem{WeW} S. Weinberg, Phys. Rev. Lett. {\bf40}(1978) 223: F. Wilczek, Phys. Rev. Lett. {\bf 40}(1978) 271.
\bibitem{mzero} D. B. Kaplan and A. V. Manohar, Phys. Rev. lett. {\bf 56}(1986) 2004.
\bibitem{Leutwyler} H. Leutwyler, Nucl. Phys. {\bf B337}(1990) 108.
\bibitem{Wolf} L. Wolfenstein, Phys. Rev. Lett. {\bf 51}(1983) 1945.
\bibitem{GW} F. Gilman and M. Wise, Phys. Lett. B{\bf 83}(1979) 83;
Phys. Rev. D{\bf 20}(1979) 83; B. Guberina and R. D. Peccei, Nucl.
Phys. B{\bf 163}(1980) 289.
\bibitem{Shabalin} E. P. Shabalin, Sov. J. Nucl. Phys. {\bf 28}(1978) 75.
\bibitem{edm} See, for example, Y. Nir in Proceedings of the 1992 SLAC
Summer School (SLAC, Stanford, California, 1993).
\bibitem{PW} R. D. Peccei and K. Wang, Phys. Lett. {\bf B349}(1995) 220. 
\bibitem{Forty} R. Forty, to appear in the Proceedings of the
International Conference on High Energy Physics (ICHEP 94), Glasgow,
Scotland, July 1994.
\bibitem{CDF} CDF Collaboration: F. Abe {\it et al.}, Phys. Rev. D{\bf 50}
(1994) 2966.
\bibitem{Stone} S. Stone, Proceedings of the 1994
DPF Conference, Albuquerque, New Mexico, ed. Sally Seidel (World Scientific, Singapore, 1995).
\bibitem{Sharpe} S. Sharpe, Nucl. Phys. B (Proc. Suppl.){\bf 34}(1994) 403.
\bibitem{Lattice} This is a guesstimate of the value of this parameter
coming from Lattice QCD calculations.  For a compilation of the most
recent results, see J. Shigemitsu, to appear in the Proceedings of
the International Conference on High Energy Physics (ICHEP 94), Glasgow,
Scotland, July 1994.
\bibitem{ALEPH} Y. B. Pan, to appear in the Proceedings of the
International Conference on High Energy Physics (ICHEP 94), Glasgow,
Scotland, July 1994.
\bibitem{E731} E731 Collaboration: L. K. Gibbons {\it et al.}, Phys.
Rev. Lett. {\bf 70}(1993) 1203.
\bibitem{NA31} NA31 Collaboration: G. D. Barr {\it et al.}, Phys. Lett.
B{\bf 317} (1993) 1233.
\bibitem{BL} A. J. Buras and M. E. Lautenbacher, Phys. Lett. B{\bf 318}
(1993) 212.  See also, A. J. Buras, in {\bf Phenomenology of Unification
from Present to Future}, ed. G. Diambrini-Palazzi {\it et al.}
(World-Scientific, Singapore, 1994).
\bibitem{Gasser} J. Gasser and U. G. Meissner, Phys. Lett. {\bf B258}(1991) 219. \bibitem{FR} J. Flynn and L. Randall, Phys. Lett. B{\bf 216}(1989) 221;
{\it ibid.} B{\bf 224}(1989) 221; Nucl. Phys. B{\bf 326}(1989) 3.
\bibitem{Ciuchini} M. Ciuchini, E. Franco, G. Martinelli, and L. Reina,
Phys. Lett. B{\bf 301}(1993) 263.
\bibitem{N} A. J. Buras, M. Jamin, and M. T. Lautenbacher, Nucl. Phys.
B{\bf 408}(1993) 209.
\bibitem{RW} For a review see, J. C. Ritchie and S. G. Wojcicki,
Rev. Mod. Phys. {\bf 65}(1993) 113.
\bibitem{Cheng} H. Y. Cheng, Phys. Rev. D{\bf 43}(1991) 1579.
\bibitem{pessimistic} J. F. Donoghue, B. R. Holstein, and G. Valencia,
Phys. Rev. D{\bf 36}(1987) 798.
\bibitem{Belkov} A. A. Belkov {\it et al.}, Phys. Lett. B{\bf 232}(1989)
118.
\bibitem{IMP} G. Isidori, L. Maiani and A. Pugliese, Nuc. Phys. B{\bf 381}
(1992) 522.
\bibitem{Pettit} These results come from the Ph.D. Thesis of F. Pettit
(UCLA 1995).  They are quite similar to those obtained by G. Isidori,
L. Maiani, and A. Pugliese, Nucl. Phys. B{\bf 381}(1992) 522.
\bibitem{HYC} H. Y. Cheng, Phys. Lett. B{\bf 315}(1993) 170;
Phys. Rev. D{\bf 49}(1994) 3771; N. Paver, Riazzudin, and F. Simeoni,
Phys. Lett. B{\bf 316}(1993) 397.  These papers supercede an earlier,
more optimistic, estimate of C. O. Dib and R. D. Peccei, Phys. Lett.
B{\bf 249} (1990) 325.
\bibitem{Seghal} L. M. Seghal, Phys. Rev. D{\bf 38}(1988) 808;
D{\bf 41}(1991) 161; T. Morozumi and H. Iwasaki, Prog. Theor. Phys.
{\bf 82}(1989) 371.
\bibitem{Dan} A. G. Cohen, G. Ecker, and A. Pich, Phys. Lett. B{\bf 304}
(1993) 347; see also G. D'Ambrosio {\it et al.}, CERN TH-7503, 1994.
\bibitem{Dib} C. O. Dib, I. Dunietz, and F. Gilman, Phys. Lett. B{\bf 218}
(1989) 487; Phys. Rev. D{\bf 39}(1989) 2639.
\bibitem{WW} See for example, B. Winstein and L. Wolfenstein, Rev.
Mod. Phys. {\bf 65}(1993) 1113.
\bibitem{BLMM} A. J. Buras, M. E. Lautenbacher, M. Misiak, and
M. M\"unz, Nucl. Phys. {\bf B423}(1994) 349.
\bibitem{Harris} E779 Collaboration: D. H. Harris {\it et al}, Phys.
Rev. Lett. {\bf 71}(1993) 3914.
\bibitem{Ohl} E845 Collaboration: K. E. Ohl {\it et al.}, Phys. Rev.
Lett. {\bf 64} (1990) 2755.
\bibitem{DHH} E799 Collaboration: D. H. Harris {\it et al}, Phys. Rev.
Lett. {\bf 71}(1993) 3918.
\bibitem{Greenlee} B. H. Greenlee, Phys. Rev. D{\bf 42}(1990) 3724.
\bibitem{WINS} B. Winstein, to appear in the Proceedings of the 23rd INS Symposium, Tokyo, Japan, 1995.
\bibitem{Littenberg} L. S. Littenberg, Phys. Rev. D{\bf 39}(1989) 3222.
\bibitem{BBB} G. Buchalla and A. J. Buras, Nucl. Phys. B{\bf 400}(1993)
225.
\bibitem{Weaver} E799 Collaboration: M. Weaver, Phys. Rev. Lett. {\bf 72}(1994) 3758.
\bibitem{Dibc}C. O. Dib, Phys. Lett. B{\bf 282}(1992) 201.
\bibitem{BB2} G. Buchalla and A. J. Buras, Nucl. Phys. B{\bf 412}(1994)
106.
\bibitem{waiting} A. J. Buras, M. E. Lautenbacher, and G. Ostermaier,
Phys. Rev. D{\bf 50} (1994) 3433.
\bibitem{GL} J. Gasser and H. Leutwyler, Phys. Rept. {\bf 87C}(1982)
77.
\bibitem{LINS} L. S. Littenberg,  to appear in the Proceedings of the 23rd INS Symposium, Tokyo, Japan, 1995.
\bibitem{Atiyah} E787 Collaboration: M. S. Atiya, Phys. Rev. Lett. 
{\bf 70}(1993) 2521; Phys. Rev. D{\bf 48}(1993) R1.
\bibitem{BS} A. Carter and A. I. Sanda, Phys. Lett. {\bf 45}(1980) 952; Phys. Rev. {\bf D23}(1981) 1567; I. I. Bigi and A. I. Sanda, Nucl. Phys. {\bf B193}(1981) 95; {\bf B281}(1987) 41. 
\bibitem{Rosner} I. Dunietz and J. L. Rosner, Phys. Rev. {\bf D34}(1986) 1404; Y. Azimov, V. A. Khoze and N. G. Uraltsev, Yad. Fyz. {\bf 45}(1987) 1412; D. Du, I. Dunietz and D. Wu, Phys. Rev. {\bf D34}(1986) 3414.
\bibitem{Bunk} For an early discussion see, for example, R. D. Peccei, in the Proceedings of the Workshop on the Experimental Program at UNK, Protvino, USSR, 1987. For a more detailed exposition see, for example, I. I. Bigi, V. A. Khoze, A. I. Sanda and N. G. Uraltsev  in
{\bf CP Violation} ed. C. Jarlskog (World Scientific, Singapore, 1989).
\bibitem{bj} For a discussion see, for example, C. Jarlskog  in
{\bf CP Violation} ed. C. Jarlskog (World Scientific, Singapore, 1989).
\bibitem{WEIN} B. Weinstein, Phys. Rev. Lett. {\bf 68}(1992) 1271.
\bibitem{Nir} There have been numerous analyses of the allowed values for the angles $\alpha, \beta$ and $\gamma$. Early studies can be found in P. Krawczyk {\it et al}, Nucl. Phys. {\bf B307}(1988) 19; C. O. Dib {\it et al}, Phys. Rev. {\bf D41}(1990) 1522; Y. Nir and H. R. Quinn, Phys. Rev. {\bf D42}(1990) 1473.
\bibitem{NQ} For a discussion, see for example, Y. Nir and H. Quinn,
in {\bf B Decays}, ed. S. Stone (World Scientific, Singapore, 1992).
\bibitem{GLP} D. London and R. D. Peccei, Phys. Lett. {\bf B223}(1989) 257; B. Grinstein, Phys. Lett. {\bf B229}(1989) 280; M. Gronau, Phys. Rev. Lett. {\bf 63}(1989) 1451; Phys. Lett. {\bf B300}(1993) 163.
\bibitem{KKPS} B. Kayser, M. Kuroda, R. D. Peccei and A. I. Sanda, Phys. Lett. {\bf B237}(1990) 508.
\bibitem{CLEO} This meausurement is discussed in S. Playfer and S. Stone, HEPSY 95-01.
\bibitem{GL2} M. Gronau and D. London, Phys. Rev. Lett. {\bf 65}(1990) 3381.
\bibitem{Quinn} H. R. Quinn and A. E. Snyder, Phys. Rev. {\bf D48}(1993) 2139.
\bibitem{GW2} M. Gronau and D. Wyler, Phys. Lett. {\bf B265}(1991) 172.
\bibitem{PecceiB} See for example, R. D. Peccei, in {\bf 25 Years of
CP Violation}, ed. J. Tran Than Van (Editions Frontiers, Gif-sur-Yvette,
1990).
\bibitem{Shap} V. A. Kuzmin, V. A. Rubakov and M. E. Shaposnikov,
Phys. Lett. {\bf 155B}(1985) 36; For a review, see for example,
M. E. Shaposnikov, Physica Scripta, {\bf T36}(1991) 183.
\bibitem{Cohen} For a discussion, see for example, M. Dine, Nucl. Phys.
B (Proc. Suppl.) {\bf 37A}(1994) 127.
\bibitem{Leurer} M. Leurer, Phys. Rev. Lett. {\bf 62} (1989) 1967.
\bibitem{Sakurai} J. J. Sakurai, Phys. Rev. {\bf 109}(1958) 980.
\bibitem{Adkins} G. S. Adkins, Phys. Rev. D{\bf 28}(1983) 2285.
\bibitem{Zhitnitski} A. R. Zhitnitski, Sov. J. Nucl. Phys. {\bf 31}(1980)
529.
\bibitem{CFK} P. Castoldi, J. M. Frere, and G. L. Kane, Phys. Rev.
D{\bf 39}(1989) 2633.
\bibitem{BG} G. Belanger and C. O. Geng, Phys. Rev. D{\bf 44}(1991)
2789.
\bibitem{Kuno} Y. Kuno, Nucl. Phys. B (Proc. Suppl.) {\bf 37A}(1994)
87.
\bibitem{Weinberg} S. Weinberg, Phys. Rev. Lett. {\bf 37}(1976) 657.
\bibitem{Higgs} R. Garisto and G. Kane, Phys. Rev. D{\bf 44}(1991) 2038;
G. Belanger and C. O. Geng, Phys. Rev. D{\bf 44} (1991) 2789.
\bibitem{Grossman} Y. Grossman, Nucl. Phys. B{\bf 426}(1994) 355.
\bibitem{Blatt} S. R. Blatt {\it et al.}, Phys. Rev. D{\bf 27}(1983) 1056.
\bibitem{CF} E. Christova and M. Fabbrichesi, Phys. Lett. B{\bf 315}
(1993) 113.
\bibitem{Hall} M. Dugan, B. Grinstein, and L. Hall, Nucl. Phys. B{\bf 255}
(1985) 413.
\bibitem{Garisto} R. Garisto, Phys. Rev. D{\bf 51}(1995) 1107.
\bibitem{PINS} R. D. Peccei, to appear in the Proceedings of the 23rd INS Symposium, Tokyo, Japan, 1995.


\end{thebibliography}
\end{document}